\documentclass[conference,a4paper]{IEEEtran}

\usepackage{multirow}
\usepackage{psfrag}
\usepackage{graphicx}
\usepackage{amsmath}
\newtheorem{theorem}{Theorem}

\newtheorem{corollary}{Corollary}

\begin{document}

\sloppy

\title{A Tight Lower Bound on the Mutual Information of a Binary and an 
       Arbitrary Finite Random Variable in Dependence of the Variational
       Distance}

\author{
  \IEEEauthorblockN{Arno G. Stefani, Johannes B. Huber}
  \IEEEauthorblockA{Institute for Information Transmission (LIT)\\
    FAU Erlangen-Nuremberg\\
    Erlangen, Germany\\
    Email: \{stefani, huber\}@LNT.de} 
  \and
  \IEEEauthorblockN{Christophe Jardin, Heinrich Sticht}
  \IEEEauthorblockA{Bioinformatics, Institute for Biochemistry\\
    FAU Erlangen-Nuremberg\\
    Erlangen, Germany\\
    Email: \{christophe.jardin, h.sticht\}@biochem.uni-erlangen.de}
}



\maketitle

\begin{abstract}
  "THIS PAPER IS ELIGIBLE FOR THE STUDENT PAPER AWARD".
  
  In this paper a numerical method is presented, which finds a lower
  bound for the mutual information between a binary and an arbitrary
  finite random variable with joint distributions that have a
  variational distance not greater than a known value to a
  known joint distribution. This lower bound can be applied to mutual
  information estimation with confidence intervals.
\end{abstract}

\section{Introduction}

A tight lower bound for the mutual information between a binary and
an arbitrary finite random variable with joint distributions that
have a variational distance not greater than a known value to a
known joint distribution can be found by minimizing over this set of joint
distributions. Unfortunately, in general this minimization problem is
hard to solve, since the mutual information is not convex in the
joint distribution.

Therefore this minimization problem is split up into two subproblems.

If the marginal probability of the binary random variable is fixed, then
the mutual information can easily be minimized over the conditional
probabilities of the second random variable, since the mutual information
is convex in the conditional probabilities \cite[Theorem 2.7.4]{Cover2006}
and the set of conditional probabilities is convex (see Theorem \ref{theorem1})
and therefore this optimization problem is convex. This
constitutes the first subproblem which can easily be solved by standard
methods for convex optimization.

In the second subproblem, having a closer look on the marginal probability
distribution of the binary random variable, one first recognizes that this
is only one-dimensional since the two probabilities have to sum up to 1.
Next, the variational distance between the joint
probabilities is greater or equal than the variational distance of the
marginal probabilities, as is shown in (\ref{eq_var}). Therefore one
can simply generate sufficiently many marginal probability distributions
equidistantly in the one dimension left, solve the first subproblem for every
of these marginal probability distributions and return the smallest mutual
information calculated that way.

In the next section the notation is fixed. In section \ref{sec_results}
the details of the method are given. In section \ref{sec_num} some numerical
examples are shown.

\section{Notational setup}
\label{sec_notation}

Let $X$, $Y$ be a pair of finite discrete random variables,
with joint probability distribution
\begin{align}
p_{XY} &{}={} \{ p_{XY}(i,j) : i=1, 2, \ldots, M_x;~j=1, 2, \ldots, M_y \}.\nonumber
\end{align}
Here $X \in \mathcal{X}$ and $Y \in \mathcal{Y}$ and it is w.l.o.g.
assumed that $\mathcal{X}=\{ 1, 2, \ldots, M_x \}$ and that $\mathcal{Y}=\{ 1, 2, \ldots, M_y \}$.
The marginal probability distributions are
$p_X = \{ p_X(i) : i=1, 2, \ldots, M_x \}$ and $p_Y = \{ p_Y(j) : j=1, 2, \ldots,
M_y \}$. They are calculated from the joint probalility distributions as usual.
The conditional probability distributions are
\begin{align}
p_{Y|X} &{}={} \{ p_{Y|X}(j | i) : i=1, 2, \ldots, M_x;~j=1, 2, \ldots, M_y \},\nonumber\\
p_{X|Y} &{}={} \{ p_{X|Y}(i | j) : i=1, 2, \ldots, M_x;~j=1, 2, \ldots, M_y \}.\nonumber
\end{align}
It is defined that $p_{Y|X}p_{X} = p_{X}p_{Y|X} = p_{X|Y}p_{Y} = p_{Y}p_{X|Y} = p_{XY}$.
The product of the marginal distributions is denoted as
\begin{align}
p_{X}p_{Y} &{}={} \{ p_{X}(i)p_{Y}(j) : i=1, 2, \ldots, M_x;~j=1, 2, \ldots, M_y \}.\nonumber
\end{align}

For any two joint probability distributions $p_{XY}$, $q_{XY}$ the relative entropy
or Kullback-Leibler distance \cite{Cover2006} is defined as
\begin{align}
\label{eq_rel_entr}
D(p_{XY} \Vert q_{XY}) &{}={} \sum\limits_{i=1}^{M_x}\sum\limits_{j=1}^{M_y} p_{XY}(i,j) \log \frac{p_{XY}(i,j)}{q_{XY}(i,j)}
\end{align}
and the mutual information between $X$ and $Y$
\cite{Cover2006} as the relative entropy between the joint probability distribution
and product of the marginal probability distributions of $X$ and $Y$
\begin{align}
\label{eq_MI}
I(X;Y) = I(p_{XY}) &{}={} D(p_{XY} \Vert p_{X}p_{Y}).
\end{align}
All $\log$s are assumed to be natural if not stated otherwise.

The variational distance between two joint probability distributions is defined as
\begin{align}
V(p_{XY}, q_{XY}) &= \left\Vert p_{XY} - q_{XY} \right\Vert_1 \nonumber\\
		  &= \sum\limits_{i=1}^{M_x}\sum\limits_{j=1}^{M_y} | p_{XY}(i,j) - q_{XY}(i,j) |, \nonumber
\end{align}
and similarly for the marginal distributions. It can be easily seen, that
$V(\cdot,\cdot) \in [0,2]$ for any two probability distributions.

\section{Results}
\label{sec_results}

First it is shown that set of all conditional probability distributions
constrained by a maximal variational distance is convex.
\begin{theorem}
\label{theorem1}
Let $p_{XY}=p_{X}p_{Y|X}$ be any fixed joint probability distribution of any two
two discrete finite random variables $X$, $Y$, let $q_X$ be any fixed probability
distribution of $X$ and let $\epsilon$ be any fixed number $\in [0,2]$.
Then the set $\mathcal{Q}=\{ q_{Y|X} ~|~ V(q_{X}q_{Y|X},p_{XY}) \leq \epsilon \}$ is convex.
\end{theorem}
\begin{IEEEproof}
Let $q_{Y|X}^1, q_{Y|X}^2$ be any two conditional probability distributions $\in Q$.
Then one only has to show that the convex combination $q_{Y|X}^\lambda=\lambda q_{Y|X}^1 +
(1-\lambda)q_{Y|X}^2$, with $\lambda \in [0,1]$ is also in $Q$. Before this is done,
it is defined that $q_{XY}^1=q_{X}^{}q_{Y|X}^1$, $q_{XY}^2=q_{X}^{}q_{Y|X}^2$ and
$q_{XY}^\lambda=\lambda q_{XY}^1 + (1-\lambda) q_{XY}^2=q_{X}^{}q_{Y|X}^\lambda$. Now,
to proof that $q_{Y|X}^\lambda \in \mathcal{Q}$, one only has to show that
$V(q_{X}^{}q_{Y|X}^\lambda,p_{XY}) \leq \epsilon$. Herefore
\begin{align}
V(q_{X}^{}q_{Y|X}^\lambda,p_{XY}) &= V(q_{XY}^\lambda,p_{XY}) \nonumber\\
			       \label{eq_normball}
			       &= \left\Vert q_{XY}^\lambda - p_{XY} \right\Vert_1 \leq \epsilon,
\end{align}
where the fact that any norm ball is convex \cite[Section 2.2.3]{Boyd2004} has
been used in (\ref{eq_normball}). Also, the further constraints implied by the
probability simplex (which is convex) are no problem since an intersection of
convex sets is always convex \cite[Section 2.3.1]{Boyd2004}.
\end{IEEEproof}
Since the empty set is convex, no restriction on
$V(p_X,q_X)$ (e.g. $V(p_X,q_X) \leq \epsilon$) is necessary.
\begin{corollary}
\label{cor1}
Let $p_{XY}$ be any fixed joint probability distribution of any two
two discrete finite random variables $X$, $Y$, let $q_X$ be any fixed probability
distribution of $X$ and let $\epsilon$ be any fixed number $\in [0,2]$. Then, the
optimization problem
\begin{align}
\min\limits_{q_{Y|X}~:~V(q_{X}q_{Y|X},p_{XY}) \leq \epsilon} I(q_{X}^{}q_{Y|X})
\end{align}
is convex.
\end{corollary}
\begin{IEEEproof}
The mutual information $I(q_{X}q_{Y|X})$ is a convex function of the conditional probabilities
$q_{Y|X}$ when $q_X$ is fixed, and the set $\{ p_{Y|X}~|~V(q_{X}^{}q_{Y|X},p_{XY}) \leq \epsilon \}$
is convex.
\end{IEEEproof}

Corollary \ref{cor1} basically says that the optimization problem given is practically
solvable. However, since it is a general convex optimization problem, it can still be
cumbersome to find a suitable algorithm with the correct parameters. Fortunately the
problem can be restated in such a way, that it can be handled by disciplined convex
programming (DCP) \cite{Grant2005}, which works perfectly well for this problem as can be
seen in section~\ref{sec_num}.

The minimization problem in Corollary \ref{cor1} can not be solved in a straightforward
manner with DCP, since this would violate the no product rule of DCP (see
(\ref{eq_rel_entr}), (\ref{eq_MI})), also there is no built function in CVX (which is the
software which implements DCP) for the mutual
information as a function of the conditional probabilities when the corresponding marginal
probability is fixed. Therefore the relative entropy, which is a built in function in CVX
and is convex in its two input arguments, is used. Then it can be seen that
\begin{align}
I(X;Y)=I(q_{X}q_{Y|X})=D(q_{X}q_{Y|X} \Vert q_{X}q_{Y}), \nonumber
\end{align}
and $q_{X}(i)q_{Y|X}(j|i)$ are affine functions of $q_{Y|X}(j|i)$ as $q_{X}(i)q_{Y}(j)
=q_{X}(i)(\sum_i q_{Y|X}(j|i)q_{X}(i))$ are. Hence, the convexity of $D(\cdot,\cdot)$ is
preserved \cite[section 2.3.2]{Boyd2004}, and it is straightforward to implement the
minimization problem in Corollary \ref{cor1} with CVX with this knowledge.

Next the second subproblem, namely the minimization of the mutual information over the
marginal probability distribution $q_X$, is solved. Herefore it is first shown that
\begin{align}
V(q_{X},p_{X}) &= \left\Vert q_{X} - p_{X} \right\Vert_1 \nonumber\\
		&= \sum\limits_{i=1}^{M_x} | q_{X}(i) - p_{X}(i) | \nonumber\\
		&= \sum\limits_{i=1}^{M_x} \left| \sum\limits_{j=1}^{M_y}(q_{XY}(i,j) - p_{XY}(i,j)) \right| \nonumber\\
		&\leq \sum\limits_{i=1}^{M_x} \sum\limits_{j=1}^{M_y} | q_{XY}(i,j) - p_{XY}(i,j) | \nonumber\\
		&= V(q_{XY},p_{XY}) \nonumber\\
		\label{eq_var}
		&\leq \epsilon.
\end{align}
Therefore only $q_X$ with $V(q_{X},p_{X}) \leq \epsilon$ have to be considered.
Until here all results are applicable to any finite $M_x$, but from here the restriction
$M_x=2$ applies.
In this case $q_X$ is one dimensional obviously, and the set of all $q_X$ is simply
$\{ q_X=\{\min(p_X(1)+\gamma,1),\max(p_X(2)-\gamma,0)\}\ |\ \gamma \in [ -\frac{\epsilon}{2},
\frac{\epsilon}{2}] \}$.
Practically, the minimization problem
\begin{align}
\min\limits_{q_{XY}~:~V(q_{XY},p_{XY}) \leq \epsilon} I(q_{XY})
\end{align}
is then simply solved by generating sufficiently many $q_X$ equidistantly in $\gamma$, solve
the optimization problem of Corollary~\ref{cor1} for every $q_X$ and return the smallest
mutual information calculated that way. Here the number of $q_X$s is considered to be
sufficient if one gets a smooth graph for the mutual information minimized over the
conditional probabilities $q_{Y|X}$ as a function of $\gamma$.

\section{Discussion}
Together with the bound on the probability of a maximal variational distance between the true
joint distribution and an empirical joint distribution (see \cite{Weissman2003}, and especially
an refinement of it which drops the dependence on the true distribution \cite[Lemma 3]{Ho2010})
the given bound can be used to construct a reasonably tight lower bound of the confidence
interval for mutual information. Such an application can be found in \cite{Stefani2013}. In
mutual information estimation with confidence intervals, the bound given is especially useful,
when the marginal probability distribuition is far from being uniform. Such a situation can
be found in \cite{Othersen2012}. In the case of two binary random variables the results seem to
coincide with lower bound of \cite{Stefani2012}.

\section{Numerical examples}
\label{sec_num}
\begin{figure}[htbp]
  \centering
  \psfrag{minMI}[c][c]{$I$}
  \psfrag{gamma}[c][c]{$\gamma$}
  \includegraphics[width=0.5\textwidth]{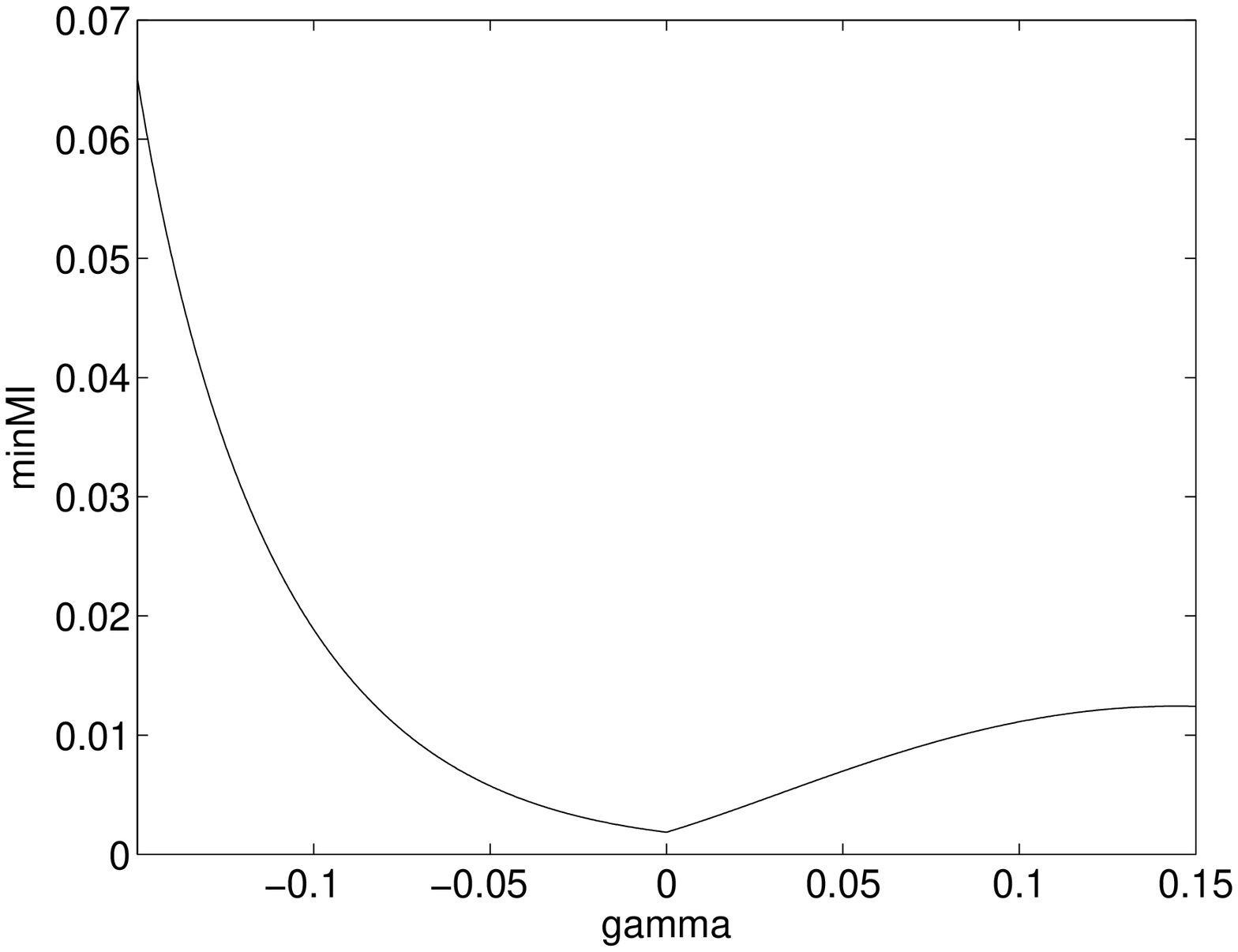}
  \caption{}
  \label{fig:MI1}
\end{figure}

In the first example (Fig.~\ref{fig:MI1}) a distribution $p_{XY}$ and a maximal variational distance $\epsilon$ was handpicked to show that the mutual information minimized over the transitional
probabilies $q_{Y|X}$ as a function of $\gamma$ is neither convex nor concave (even for two binary random variables) and seems to be not
differentiable at $\gamma=0$, as can be seen in Fig.~\ref{fig:MI1}. The parameters chosen therefore are
\begin{align*}
&p_{XY}(1,1)=0.017, p_{XY}(1,2)=0.285 \\
&p_{XY}(2,1)=0.424, p_{XY}(2,2)=0.274 \\
&\text{and }\epsilon=0.3.
\end{align*}
Then,
\begin{align*}
&I(p_{XY}) \approx 0.2210\text{ and}\\
&\min\limits_{q_{XY}~:~V(q_{XY},p_{XY}) \leq \epsilon} I(q_{XY}) \approx 0.0019.
\end{align*}

In all figures $I$ is equal to the minimum of $I(q_{X}q_{Y|X})$ over $q_{Y|X}$ for fixed
$q_X=\{\min(p_X(1)+\gamma,1),\max(p_X(2)-\gamma,0)\}$, constrained by $V(q_{X}q_{Y|X}, p_{XY})
\leq \epsilon$, and $1000$ points were generated equidistantly for $\gamma \in 
[-\frac{\epsilon}{2}, \frac{\epsilon}{2}]$.

\begin{figure}[htbp]
  \centering
  \psfrag{minMI}[c][c]{$I$}
  \psfrag{gamma}[c][c]{$\gamma$}
  \includegraphics[width=0.5\textwidth]{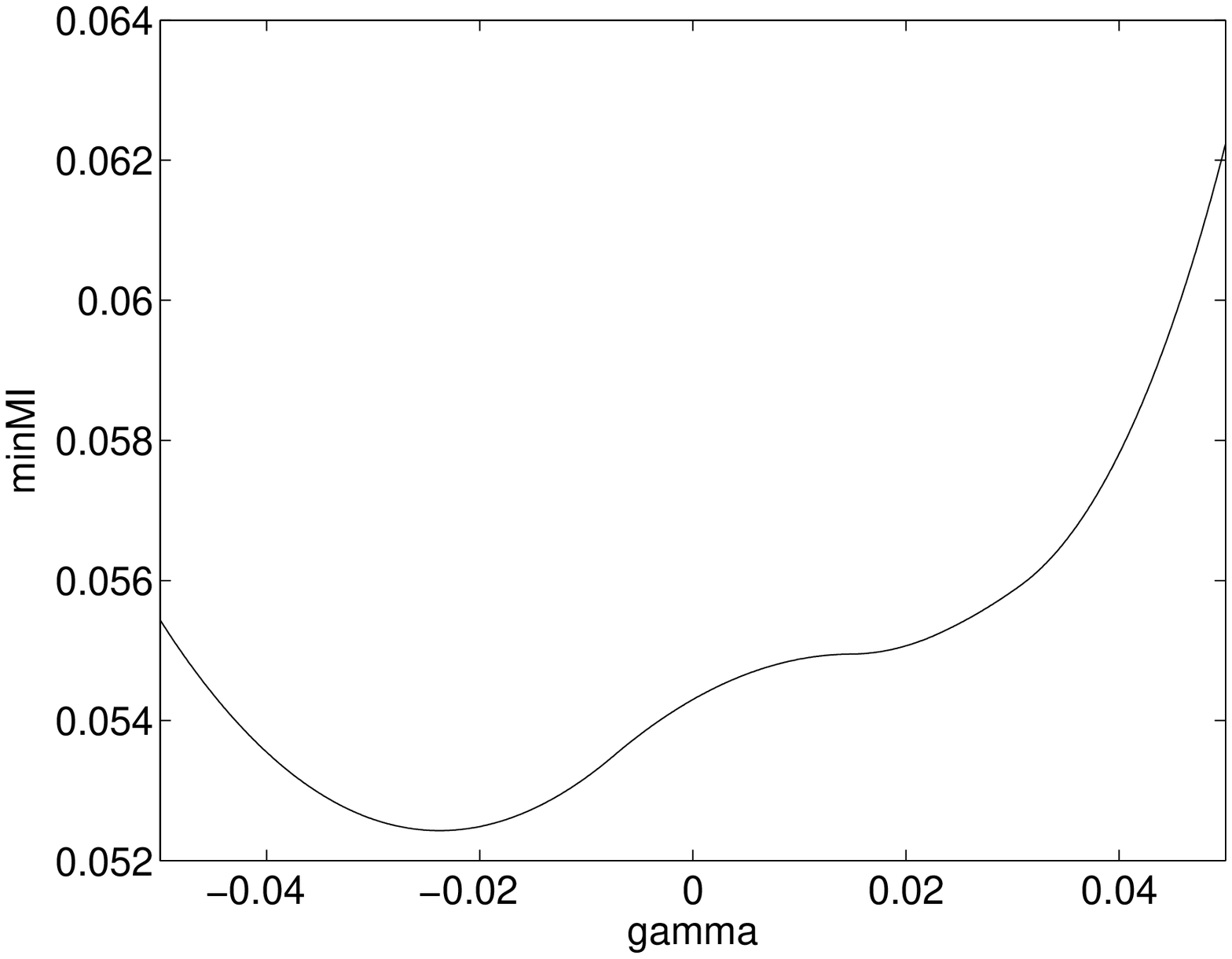}
  \caption{}
  \label{fig:MI2}
\end{figure}

In the second example (Fig.~\ref{fig:MI2}) $M_y=5$ and the following joint distribution was chosen
at random (rounded for easier reproducibility)
\begin{align*}
&p_{XY}(1,1)=0.090, p_{XY}(1,2)=0.098, p_{XY}(1,3)=0.207,\\
&p_{XY}(1,4)=0.064, p_{XY}(1,5)=0.026,\\
&p_{XY}(2,1)=0.239, p_{XY}(2,2)=0.030, p_{XY}(2,3)=0.104,\\
&p_{XY}(2,4)=0.107, p_{XY}(2,5)=0.035,\\
&\text{and } \epsilon=0.1.
\end{align*}
Then,
\begin{align*}
&I(p_{XY}) \approx 0.1112\text{ and}\\
&\min\limits_{q_{XY}~:~V(q_{XY},p_{XY}) \leq \epsilon} I(q_{XY}) \approx 0.0524.
\end{align*}

\begin{figure}[htbp]
  \centering
  \psfrag{minMI}[c][c]{$I$}
  \psfrag{gamma}[c][c]{$\gamma$}
  \includegraphics[width=0.5\textwidth]{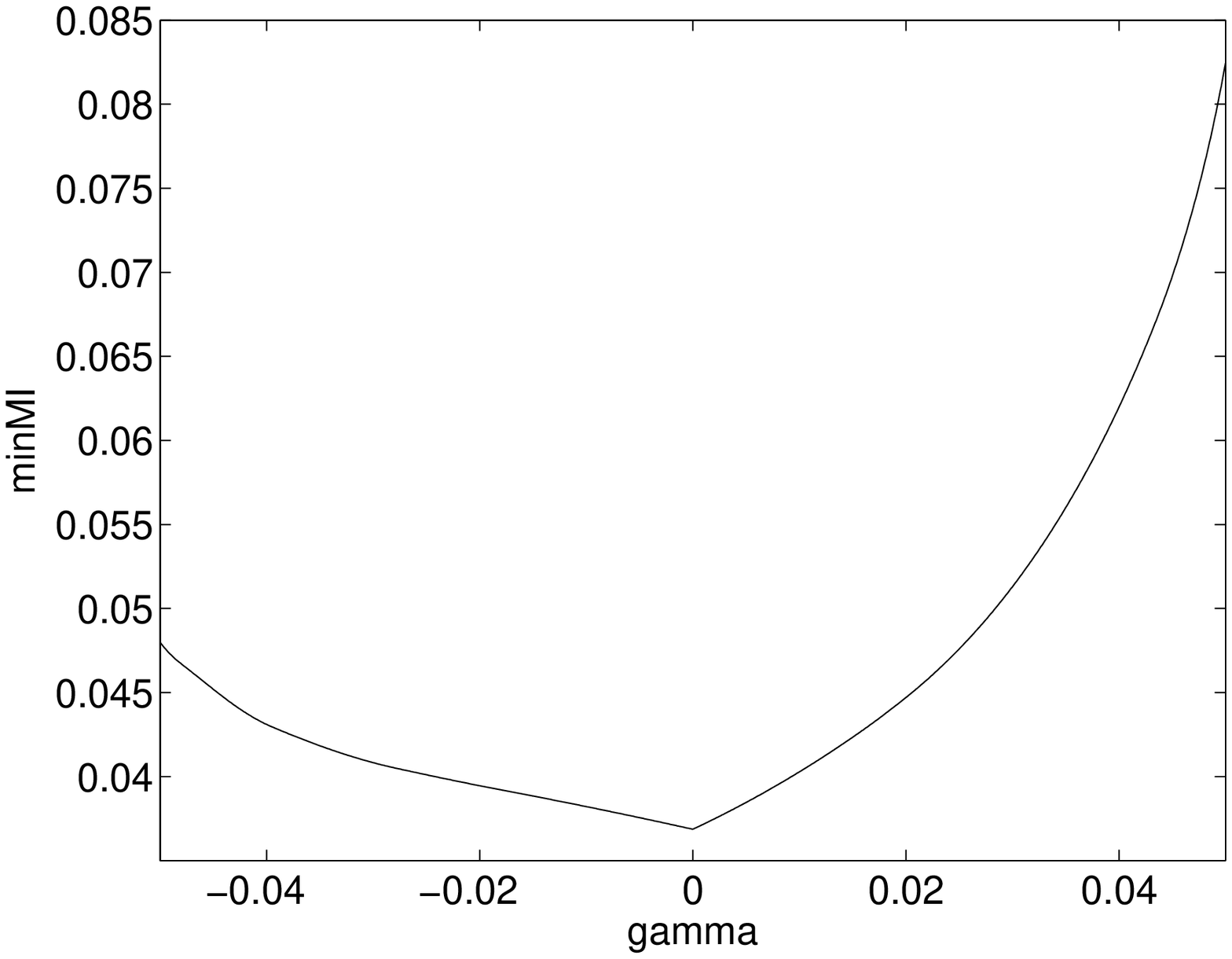}
  \caption{}
  \label{fig:MI3}
\end{figure}

In the last example (Fig.~\ref{fig:MI3}) $M_y=10$ and the following joint distribution was chosen
at random (rounded for easier reproducibility)
\begin{align*}
&p_{XY}(1,1)=0.101, p_{XY}(1,2)=0.062, p_{XY}(1,3)=0.025,\\
&p_{XY}(1,4)=0.088, p_{XY}(1,5)=0.005, p_{XY}(1,6)=0.007,\\
&p_{XY}(1,7)=0.069, p_{XY}(1,8)=0.059, p_{XY}(1,9)=0.080,\\
&p_{XY}(1,10)=0.074,\\
&p_{XY}(2,1)=0.103, p_{XY}(2,2)=0.006, p_{XY}(2,3)=0.038,\\
&p_{XY}(2,4)=0.002, p_{XY}(2,5)=0.018, p_{XY}(2,6)=0.079,\\
&p_{XY}(2,7)=0.049, p_{XY}(2,8)=0.032, p_{XY}(2,9)=0.020,\\
&p_{XY}(2,10)=0.020,\\
&\text{and } \epsilon=0.1.
\end{align*}
Then,
\begin{align*}
&I(p_{XY}) \approx 0.1311\text{ and}\\
&\min\limits_{q_{XY}~:~V(q_{XY},p_{XY}) \leq \epsilon} I(q_{XY}) \approx 0.0369.
\end{align*}
\section*{Acknowledgment}

The authors would like to thank the DFG for supporting 
their research with SPP1395 in the projects HU634\_7 and STI155\_3.


\end{document}